\documentclass[aps,prl,twocolumn,showpacs,superscriptaddress,amsmath,amssymb,groupedaddress]{revtex4-1}
\usepackage{graphicx}
\usepackage{latexsym}
\usepackage{amsmath,amssymb}
\usepackage{bm} 
\usepackage{color}
\usepackage{epsfig}

\usepackage{graphicx}
\usepackage[colorlinks=true,linkcolor=blue,citecolor=blue,urlcolor=blue]{hyperref}
\usepackage{bbold}
\usepackage{gensymb}

\usepackage{pgfplots,pgfplotstable}
\tikzset{every axis plot post/.append style={solid, thin},every mark/.append style={solid,scale=1}}
\usetikzlibrary{decorations.markings}
\tikzset{mdar/.style ={decoration={markings,mark={at position 0.5 with {\fill (2pt,0)--(-2pt,2pt)--(-2pt,-2pt)--cycle;}}},postaction={decorate}}}
\tikzstyle{dsh}=[dash pattern=on 2.5pt off 1.5pt]
\tikzset{intcur/.style ={decorate,decoration={snake,amplitude=.5mm,segment length=3mm}}}
\tikzstyle{zba}=[dash pattern=on 1pt off 1pt,line width=2pt]

\usepackage{simplewick}

\definecolor{myblue}{rgb}{.93, .93, 1}

\def \a {\alpha}

\def \D {\Delta}

\def \G{\Gamma}

\def \dag {\dagger}

\def \grad {\nabla}

\def \eqv {\equiv}
\def \apx {\approx}

\def \dag {\dagger}

\newcommand{\intv}[1]{\int_{\mbf #1}}

\def \la {\langle}
\def \ra {\rangle}

\newcommand{\ket}[1]{|#1\ra}
\newcommand{\braket}[3]{\la#1|#2|#3\ra}

\newcommand{\epvl}[1]{\la#1\ra}

\def \Tr {\mathrm{Tr}}
\def \bece {\begin{center}}
\def \ence {\end{center}}
\def \beeq {\begin{equation}}
\def \eneq {\end{equation}}
\def \beal {\begin{aligned}}
\def \enal {\end{aligned}}
\def \bega {\begin{gathered}}
\def \enga {\end{gathered}}
\def \benu {\begin{enumerate}}
\def \ennu {\end{enumerate}}
\def \beit {\begin{itemize}}
\def \enit {\end{itemize}}
\def \bede {\begin{description}}
\def \ende {\end{description}}
\def \betb {\begin{tabular}}
\def \entb {\end{tabular}}
\def \bear {\begin{array}}
\def \enar {\end{array}}

\def \mbf {\mathbf}
\def \mbb {\mathbb}

\def \bsb{\boldsymbol}
\def \txt {\text}

\newcommand{\comment}[1]{}

\begin{document}


\title{Higher-order topological insulators from $3Q$ charge bond orders on hexagonal lattices: A hint to kagome metals}

\author{Yu-Ping Lin}
\affiliation{Department of Physics, University of Colorado, Boulder, Colorado 80309, USA}

\date{\today}

\begin{abstract}
We show that unconventional boundary phenomena occur in the $3Q$ charge bond orders on the hexagonal lattices. At the Van Hove singularity with three nesting momenta, $3Q$ orders can trigger a $\text{C}_6$-symmetric insulator under bond modulations. On the kagome lattice, in-gap corner states appear in the energy spectrum and carry fractional corner charge $-2e/3$. Such corner phenomena originate from the corner filling anomaly and indicate a higher-order topological insulator. The in-gap corner states are also observed on the triangular lattice. The honeycomb lattice does not support fractional corner charges, while in-gap edge states are observed. We discuss possible indications to the experimentally uncovered charge bond orders in the kagome metals $\text{AV}_3\text{Sb}_5$ with $\text{A}=\text{K},\text{Rb},\text{Cs}$. With layer stacking along the out-of-plane direction, the corner states can constitute the hinge states with fractional charge densities.
\end{abstract}

\maketitle


Recent decades have witnessed advanced understanding of condensed matter phenomena under symmetries \cite{hasan10rmp}. Particular efforts have been devoted to the crystalline symmetries, where various topological crystalline insulators are uncovered \cite{fu10prl,ando15tci}. Boundary anomalies lie at the heart of these topological phases, where fractional charges may arise from the bulk topologies. In the past few years, lower-dimensional boundary anomalies have been identified in a new class of higher-order topological insulators \cite{benalcazar17sc,benalcazar17prb,benalcazar19prb,schindler18sc,schindler18np,langbehn17prl,geier18prb,trifunovic19prx,song17prl,khalaf18prb,trifunovic21pssb}. Paradigmatic examples include the second-order topological insulators under $\txt{C}_n$ symmetry in two dimensions. While the bulk and edges are gapped, corner states with fractional charges arise under corner filling anomaly \cite{benalcazar19prb,peterson20sc}. Notably, the corner states can also appear in the other situations \cite{wang19prl,ahn19prx,benalcazar19prb,khalaf21prr,ezawa18prl,ni19nm,kempkes19nm}. Several methods may serve as complementary examinations of higher-order topology, including the computations of Wilson loop \cite{benalcazar17sc,benalcazar17prb}, entanglement spectrum \cite{zhu20prb}, and $\mbb Z_Q$ Berry phase \cite{araki20prr}.

\begin{figure}[t]
\centering
\includegraphics[scale = 1]{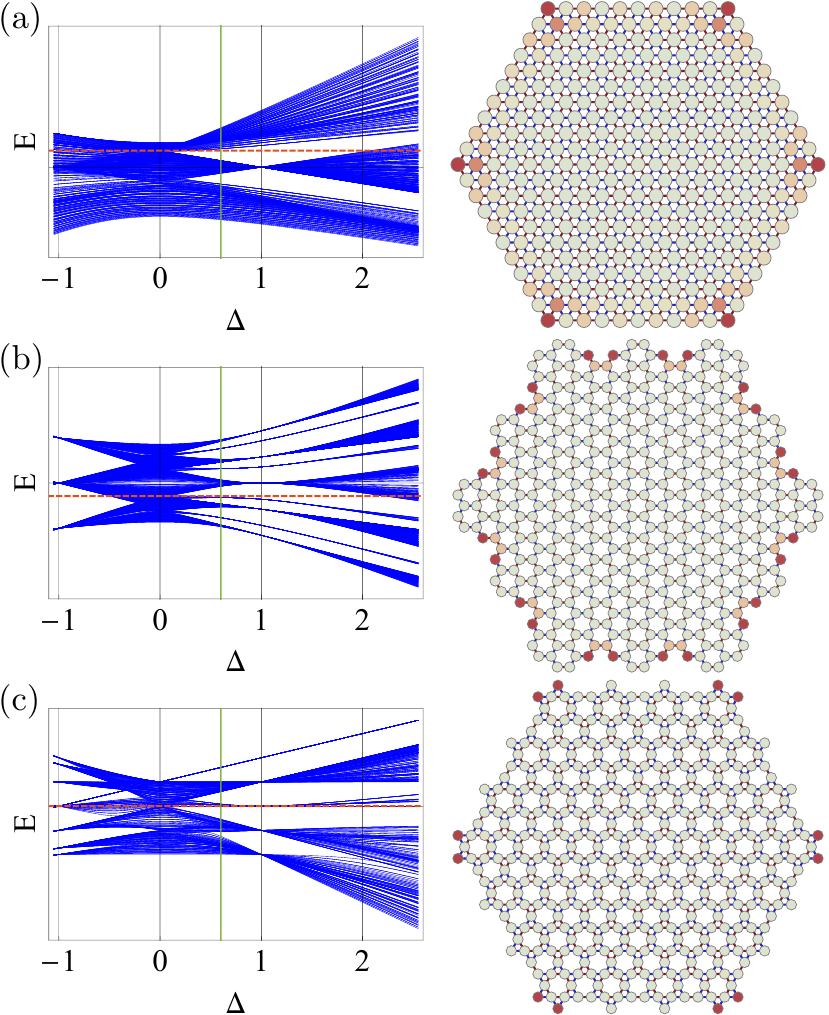}
\caption{\label{fig:es} (Left) Energy spectra under $3Q$ charge bond orders $\D$. The in-gap states on the (a) triangular lattice tie closely to the bottom of unoccupied bulk band, while clear in-gap branches appear on the (b) honeycomb and (c) kagome lattices. (Right) Real-space density patterns of the eigenstates closest to the Fermi level (left red dashed lines) at $\D=0.6$ (left green lines). The kagome and triangular lattices present corner states, while the honeycomb lattice hosts edge states. Note that the corner states on the kagome lattice almost overlap with the Fermi level below the low-energy curves.}
\end{figure}

Most proposals to higher-order topology address noninteracting solid-state systems or artificial models. Meanwhile, the practical solid-state materials manifest inevitable interactions, thereby hosting various symmetry-breaking phases. These symmetry-breaking phases, such as superconductors \cite{wang18prb,hsu20prl,zhang20prb}, can offer important platforms for the practical realization of higher-order topology. The family of hexagonal lattices, including triangular, honeycomb, and kagome lattices, have served as fertile grounds for symmetry-breaking phases. Particular interests have surrounded the saddle point of dispersion energy, where the density of states experiences the logarithmic Van Hove singularity \cite{vanhove53pr}. Various phases have been proposed at this point under strongly amplified interactions \cite{martin08prl,nandkishore12np,nandkishore12prl,kiesl12prb,lin19prb,yu12prb,kiesel13prl,wang13prb,venderbos16prbcdw,lin21ax}. Recently, a new class of kagome metals $\txt{AV}_3\txt{Sb}_5$, where $\txt{A}=\txt{K},\txt{Rb},\txt{Cs}$, have been observed to host unconventional charge density waves at Van Hove singularity \cite{yang20sa,jiang20ax,yu21ax,ortiz20prl,zhao21ax,uykur21ax,wang21ax,li21ax,shumiya21ax,wang21axpu,kang21ax}. These states develop at $80$-$110$ $\txt{K}$ far above the superconductivity at $0.9$-$2.7$ $\txt{K}$. The $3Q$ charge bond orders at three nesting momenta are likely manifest, where commensurate bond modulations form ``inverse star-of-David" patterns on the $\txt{V}$ kagome layers \cite{tan21ax,feng21scb,denner21ax,lin21ax,park21ax,setty21ax,feng21ax,miao21ax}. Numerous unconventional phenomena have been uncovered, including the momentum-dependent gaps and giant anomalous Hall effect. The $3Q$ charge bond orders are usually identified as trivial insulators with zero Chern number \cite{venderbos16prbcdw,lin21ax}. With the precedent example of Kekul\'e order at the Dirac points \cite{benalcazar19prb,liu19prl,fleury19prl,lee20prb,mizoguchi19jpsj}, it is tempting to ask whether higher-order topology also adds a new color to these kagome metals $\txt{AV}_3\txt{Sb}_5$.

Motivated by the observations in the kagome metals $\txt{AV}_3\txt{Sb}_5$, in this Letter we examine whether unconventional boundary phenomena occur in the $3Q$ charge bond orders on the hexagonal lattices. The analysis offers evident indications to the higher-order topology. Our work focuses on the $\txt{C}_6$-symmetric insulators under bond modulations. On the kagome lattice [Fig.~\ref{fig:es}(c)], in-gap corner states appear in the energy spectrum and carry fractional corner charge $-2e/3$. Such corner phenomena originate from the corner filling anomaly and indicate a higher-order topological insulator \cite{benalcazar19prb}. The in-gap corner states are also observed on the triangular lattice [Fig.~\ref{fig:es}(a)]. The honeycomb lattice does not support fractional corner charges, while in-gap edge states are observed [Fig.~\ref{fig:es}(b)]. We discuss possible indications to the experimentally uncovered charge bond orders in the kagome metals $\txt{AV}_3\txt{Sb}_5$. With layer stacking along the out-of-plane direction, the corner states can constitute the hinge states with fractional charge densities.


\begin{figure}[b]
\centering
\includegraphics[scale = 1]{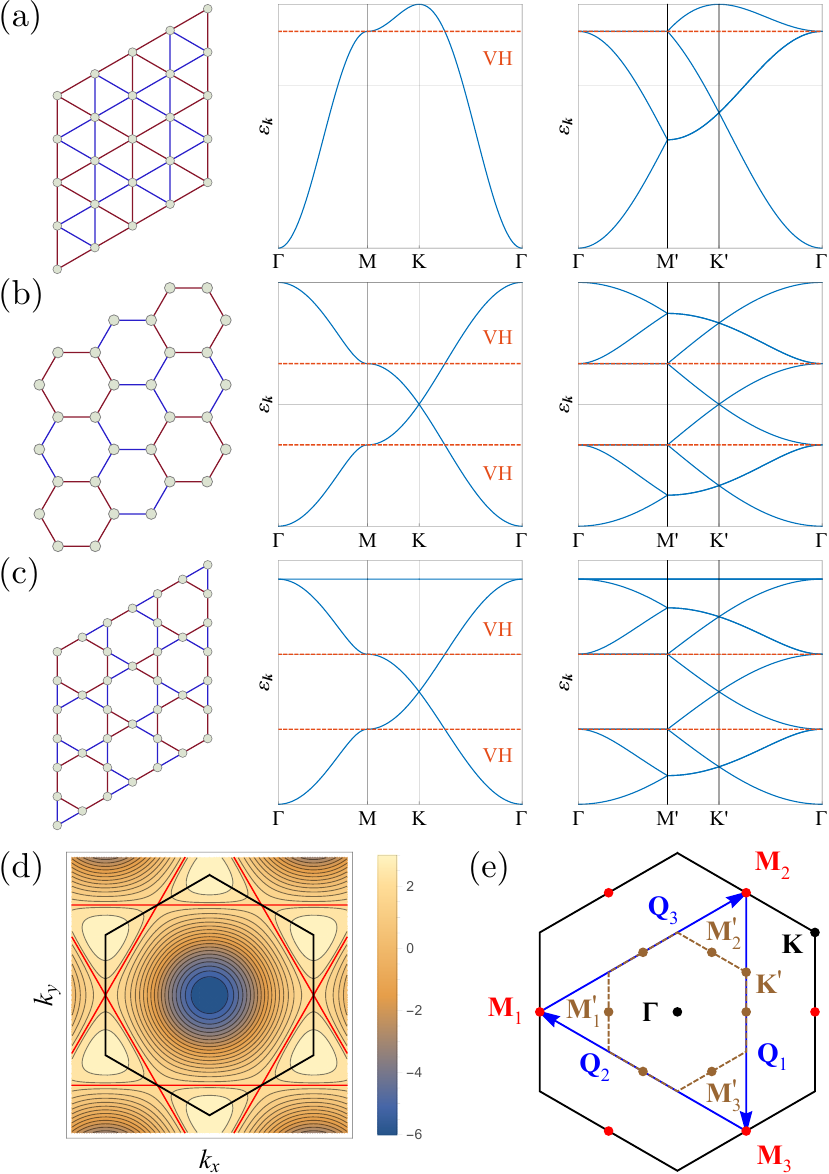}
\caption{\label{fig:lattices} For the (a) triangular, (b) honeycomb, and (c) kagome lattices, we present the (left) lattices, (middle) nearest-neighbor tight-binding band structures with Van Hove singularity (VH, red dashed lines), and (right) folded bands in the reduced Brillouin zone. The bond modulations of $3Q$ charge bond orders are shown on the lattices, where the warm and cold colors indicate strong and weak bonds, respectively. (d) Unfolded band structure. The hexagonal Fermi surface (red) is inscribed in the hexagonal Brillouin zone (black). (e) High-symmetry points in the Brillouin zone. The saddle points $\mbf M_\a$ are connected by the nesting momenta $\mbf Q_\a$. The inner hexagon indicates the reduced Brillouin zone.}
\end{figure}

Our point of departure is the Van Hove singularity on the hexagonal lattices, including triangular, honeycomb, and kagome lattices (Fig.~\ref{fig:lattices}). The Van Hove Fermiology is universal, where a hexagonal Fermi surface exhibits corner saddle points at the edge centers $\mbf M_{\a=1,2,3}$ of hexagonal Brillouin zone. The parallel Fermi lines manifest opposite energy structures, thereby supporting strong Fermi surface nesting at three momenta $\mbf Q_\a\eqv\mbf M_\a$. Combining Van Hove singularity and Fermi surface nesting, the strongly amplified interactions can trigger charge bond orders at the nesting momenta $\mbf Q_\a$ \cite{park21ax}. The $3Q$ orders are energetically favored, where equal-strength orders develop simultaneously at three momenta $\mbf Q_\a$ \cite{lin21ax}. Under the commensurate orders with $2\mbf Q_\a\eqv0$, the lattice periodicity is doubled. The unit cell is enlarged to a $2\times2$ structure, accompanied by a $1/2\times1/2$ reduced Brillouin zone. The bands are folded onto the reduced Brillouin zone and become quadrupled. These bands are then modified by the charge bond orders, where a full gap can be opened at the Fermi level \cite{venderbos16prbcdw,lin21ax}.

\begin{figure*}[t]
\centering
\includegraphics[scale = 1]{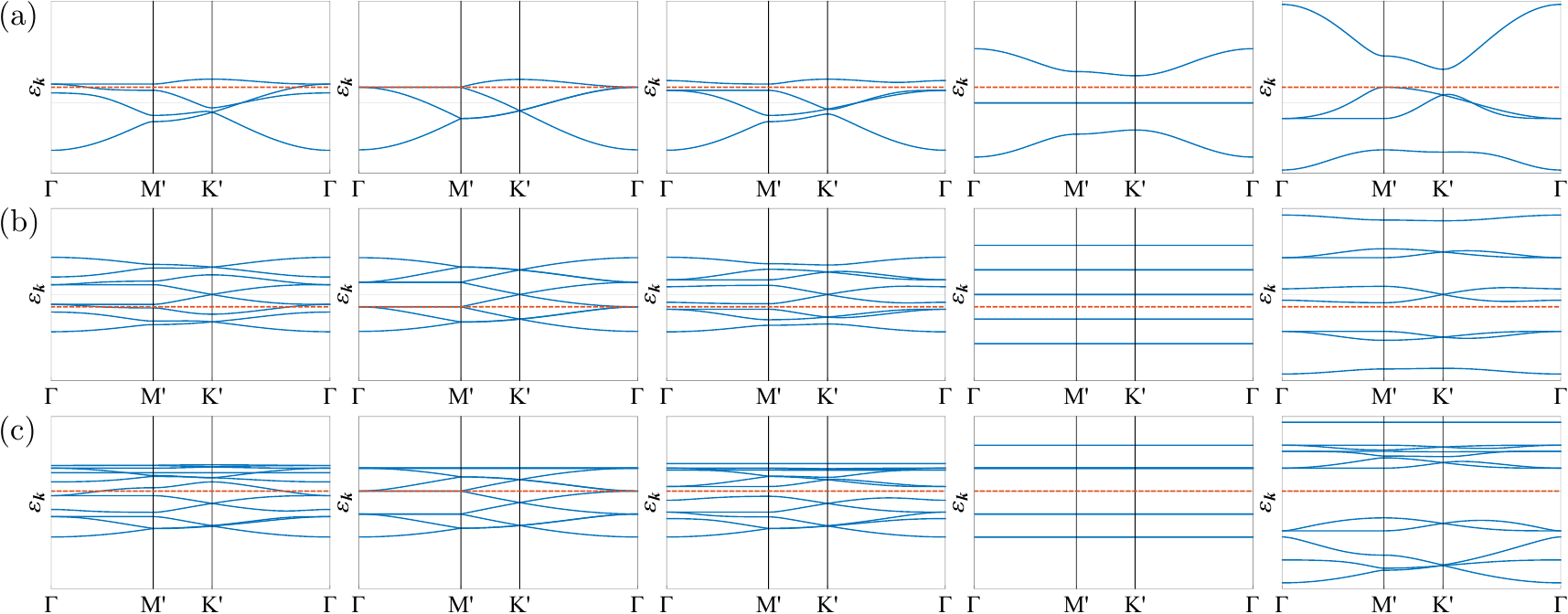}
\caption{\label{fig:bands} The band structures under $3Q$ charge bond orders on the (a) triangular, (b) honeycomb, and (c) kagome lattices. We choose the orders $\D=-0.2$, $0$, $0.2$, $1$, $2$ from the left to the right. The system becomes fully gapped at the Fermi level (red dashed lines) under $\D>0$. Flat bands occur at the exactly solvable point $\D=1$ due to the vanishing of weak bonds $t_-=0$. Meanwhile, the Fermi surface remains finite under small $\D<0$.}
\end{figure*}

We adopt an effective tight-binding model
\beeq
H=-\sum_{\epvl{ij}_1}t_\pm c_i^\dag c_j+\txt{H.c.}
\eneq
to describe the mean-field theory of $3Q$ charge bond orders. The bond modulations are captured by the strong and weak bonds $t_\pm=t\pm\D$, where $t=1$ defines the nearest-neighbor hoppings of fermions $c_i$ and $\D$ characterizes the bond orders. In the absence of bond orders $\D=0$, the Fermi surface exhibits three $\mbf M_\a$-$(-\mbf M_\a)$ lines crossing at the Brillouin zone center $\bsb\G$. This Fermi surface can be gapped at finite orders $\D\neq0$ \cite{venderbos16prbcdw,lin21ax}. We define the convention where the full gap occurs at $\D>0$ (Fig.~\ref{fig:bands}). This gapped phase will be the regime of our focus. Note that the interaction-driven orders only explore the perturbative regime $|\D|\ll t$. Nevertheless, we extend the analysis to the strong-order regime $\D\apx t$, which will prove advantageous in later analysis. A particular benefit is the reach to an exactly solvable point $\D=1$ with vanishing weak bonds $t_-=0$. Many ground state properties can be addressed transparently at this point. On the other hand, the Fermi surface remains finite under small orders $\D<0$, while a full gap may be opened at larger orders.

With the gapped phase identified from the $3Q$ charge bond orders, a natural question is what ground state properties are manifest within this phase. The ground state properties depend strongly on the underlying symmetries. Since the charge bond orders are real, time-reversal symmetry is preserved upon ordering. This implies a zero total Chern number for the occupied bands \cite{venderbos16prbcdw,lin21ax}. Meanwhile, the bond orders break the 1-site translation symmetries and result in a $\mbb Z_4$ ground state manifold \cite{nandkishore12prl}. Rotation symmetries serve as important indicators to the topological properties. Due to the commensuration of momenta $\mbf Q_\a$, the charge bond orders are invariant under inversion $\txt{C}_2$ symmetry. The $3Q$ orders further preserve the $\txt{C}_3$ symmetry, which leads to an overall $\txt{C}_6$ symmetry. An important topological index under discrete rotation symmetries is the polarization $\mbf p=(1/S)\intv{k}\Tr\mbf A_{\mbf k}$ \cite{benalcazar17prb}, where $S$ is the Brillouin zone volume and $\mbf A^{mn}_{\mbf k}=\braket{u^m_{\mbf k}}{i\grad_{\mbf k}}{u^n_{\mbf k}}$ is the Berry connection of band eigenstates $\ket{u^n_{\mbf k}}$. Since the polarization vanishes under $\txt{C}_6$ symmetry \cite{benalcazar19prb}, the edge anomalies do not occur. Nevertheless, unconventional phenomena may still occur at the corners. The examination of boundary phenomena will be the main task of our following investigations.

\begin{figure}[t]
\centering
\includegraphics[scale = 1]{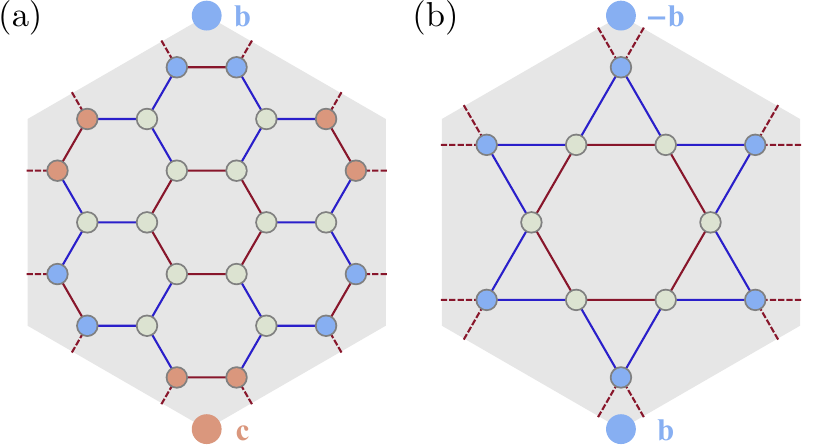}
\caption{\label{fig:uc} The hexagonal unit cells on the (a) honeycomb and (b) kagome lattices. Following the symmetry classification \cite{benalcazar19prb}, the unit cells can be decomposed into the primitive generators of topological crystalline insulators. The honeycomb unit cell is composed of two primitive generators $h^{(3)}_{(6,3)b\txt{/}c}$ (blue and red) and trivial insulators (green). Meanwhile, the kagome unit cell is composed of $h^{(6)}_{(6,2)b}$ (blue) and trivial insulator (green). The corner dots indicate the maximal Wyckoff positions, which are the locations of Wannier centers in the primitive generators.}
\end{figure}

The adoption of finite-size systems is necessary for the boundary analysis. In the formal discussion of boundary anomaly under $\txt{C}_n$ symmetry, a unit cell compatible with $\txt{C}_n$ symmetry is always chosen \cite{benalcazar19prb}. An open boundary is then determined without cutting through any unit cell. The bulk-boundary correspondence can be established in this scenario, where the bulk topology is reflected by the boundary anomaly. However, an incompatibility between the unit cell and $\txt{C}_n$ symmetry may occur in the density wave orders. This happens particularly in the $3Q$ charge bond order on the triangular lattice, where the $4m$-site unit cells are incompatible with $\txt{C}_6$ symmetry [which needs $(6m+1)$-site unit cells]. We will adopt a hexagonal finite-size system respecting $\txt{C}_6$ symmetry, with the subtlety in mind that the open boundary cuts through some unit cells. On the other hand, the $3Q$ charge bond orders exhibit $8m$- and $12m$-site unit cells on the honeycomb and kagome lattices, respectively. The proper unit cells under $\txt{C}_6$ symmetry can be safely chosen (Fig.~\ref{fig:uc}), and the hexagonal open boundaries can be defined accordingly.

We first examine the boundary phenomena of $3Q$ charge bond order on the triangular lattice [Fig.~\ref{fig:es}(a)]. The finite-size system is placed on a hexagon with strong-bond boundary. An exact diagonalization yields the evolution of energy spectrum under the bond order $\D$. The opening of a full gap at the Fermi level is clearly observed at $\D>0$. Remarkably, in-gap boundary states appear at the bottom of unoccupied bulk band. These states are well localized at the corners, indicating their nature of corner states. Similar results are also uncovered on the kagome lattice [Fig.~\ref{fig:es}(c)]. Significantly, the in-gap corner states now appear at nearly zero energies, which suggests the possibility of experimental observations. On the other hand, the honeycomb lattice does not host in-gap states at the corners [Fig.~\ref{fig:es}(b)]. Instead, the in-gap states are well localized along the edges.

We have observed the corner and edge states in the $3Q$ charge bond orders. An essential question is whether these states originate from nontrivial bulk topology. A possible origin of corner phenomena is the corner filling anomaly, which can occur under vanishing bulk polarization \cite{benalcazar19prb}. The central concept is that the charge neutrality may become incompatible with $\txt{C}_n$ symmetry under nontrivial bulk topology. The charge neutrality in an insulator is achieved by equal number of electrons and ions with charges $\mp e$. Under $\txt{C}_n$ symmetry, the charge centers sit at the high-symmetry points in each unit cell, known as the maximal Wyckoff positions. The ionic charge center sit at the unit cell center $a$. Meanwhile, the electronic charge centers can be shifted by nontrivial bulk topology. In the obstructed atomic insulators with Wannier representations, the electronic charge centers are the shifted Wannier centers at the edge centers or corners. When an open boundary is created, these Wannier centers are distributed asymmetrically on the boundary. Additional corner electrons or holes are supplemented to fulfill $\txt{C}_n$ symmetry, thereby causing  the breakdown of charge neutrality. Under this corner filling anomaly, fractional corner charges can appear and serve as strong evidences of higher-order topological insulators.

For the $3Q$ charge bond order on the kagome lattice, the corner phenomena hold a clear origin from the corner filling anomaly. Following the symmetry classification \cite{benalcazar19prb}, we identify each unit cell as the addition of a $6$-site primitive generator $h^{(6)}_{(6,2)b}$ (corresponds to $h^{(6)}_{4b}$ in the symmetry classification \cite{benalcazar19prb} with opposite sign of hoppings) and a $6$-site trivial insulator [Fig.~\ref{fig:uc}(b)]. These two elements are completely decoupled at the exactly solvable point $\D=1$. Under $\txt{C}_6$ symmetry, the primitive generator $h^{(6)}_{(6,2)b}$ fills 2 out of 6 bands by occupying the maximal Wyckoff positions $\pm b$ at the corners. According to the corner filling anomaly, each $2\pi/6$ angle sector exhibits a total fractional corner charge $-2e/3$. Note that the corner states are not necessarily robust zero modes at the Fermi level \cite{hughes11prb}. Although a generalized chiral symmetry is present in the bulk \cite{ni19nm,kempkes19nm}, there is no particle-hole or chiral symmetry on the edges which protects the zero modes. All of the ground states in the gapped phase can be continuously connected to the exactly solvable point $\D=1$. We thus conclude that the gapped phase of $3Q$ charge bond order is a higher-order topological insulator on the kagome lattice. Similar analysis applies to the honeycomb lattice, which is the addition of two $6$-site primitive generators $h^{(3)}_{(6,3)b\txt{/}c}$ and two $6$-site trivial insulators [Fig.~\ref{fig:uc}(a)]. While the primitive generators $h^{(3)}_{(6,3)b\txt{/}c}$ respect the lower $\txt{C}_3$ symmetry, their addition manifests a $\txt{C}_6$-symmetric insulator with identical Wannier centers to those in $h^{(6)}_{(6,2)b}$. However, each Wannier center now carries a charge $-3e$, leading to an integer charge at each corner. Therefore, the $3Q$ charge bond order does not host fractional corner charges on the honeycomb lattice. Since the edge anomaly is absent under vanishing bulk polarization, the observed edge states may arise from some nonanomalous origin. The corner states are also observed on the triangular lattice, despite the subtlety in the cut of boundary unit cells. The understanding of this corner phenomena may be an interesting topic for future work.

We have identified the higher-order topology of $3Q$ charge bond order on the kagome lattice. A potential platform for its realization is the kagome metals $\txt{AV}_3\txt{Sb}_5$ with $\txt{A}=\txt{K},\txt{Rb},\txt{Cs}$. The experiments have observed unconventional charge density waves on the $\txt{V}$ kagome layers \cite{yang20sa,jiang20ax,yu21ax,ortiz20prl,zhao21ax,uykur21ax,shumiya21ax,li21ax,wang21ax,wang21axpu,kang21ax}, which seemingly manifest 3Q charge bond orders at Van Hove singularity \cite{tan21ax,feng21scb,denner21ax,lin21ax,park21ax,setty21ax,feng21ax,miao21ax}. Based on our analysis, we propose that these states realize higher-order topological insulators on the $\txt{V}$ kagome layers. Corner states can arise at low energy, which carry fractional corner charge $-2e/3$ under corner filling anomaly. With layer stacking along the out-of-plane direction, the corner states can assemble into the hinge states with fractional charge densities. A caveat is that $\txt{C}_6$ symmetry breaking has been widely observed in the experiments, where different strengths of bond modulations occur at different momenta. Such observations may be explained by an interlayer effect, where adjacent layers manifest different $3Q$ orders in the $\mbb Z_4$ ground state manifold \cite{park21ax,miao21ax}. Since $\txt{C}_6$ symmetry may persist on each $\txt{V}$ kagome layer in this scenario, the higher-order topological insulators may be protected. If the intralayer $\txt{C}_6$ symmetry were broken, $\txt{C}_2$ symmetry might still preserve the hinge states with fractional charge density $-e/2$ in each $2\pi/2$ angle sector \cite{benalcazar19prb}. Whether the hinge states exist in the practical systems is an interesting topic for future experiments. On the other hand, the experiments suggest the presence of residual Fermi surfaces from finite doping, nonperfect nesting, or the other bands. Furthermore, the giant anomalous Hall effect indicates time-reversal symmetry breaking, which may be attributed to the onset of flux orders \cite{venderbos16prbcdw,lin19prb,feng21scb,denner21ax,lin21ax,park21ax,feng21ax}. The robustness of hinge states in these situations is an important issue for future study. Finally, the $3Q$ charge bond orders may also be simulated in the ultracold atomic systems \cite{cooper19rmp}. The realization of higher-order topology is an interesting topic for experimental investigations.


In summary, we have uncovered higher-order topological insulators from the $3Q$ charge bond orders on the hexagonal lattices. Under $\txt{C}_6$ symmetry, corner filling anomaly brings fractional corner charges to the in-gap corner states. With the enhancement from Van Hove singularity, such a topological phase may arise in a variety of materials. Potential candidates include the recently found kagome metals $\txt{AV}_3\txt{Sb}_5$ with $\txt{A}=\txt{K},\txt{Rb},\txt{Cs}$, where hinge states may emerge from layer stacking and carry fractional charge densities. Our work sheds light on the search for higher-order topological phases in the practical materials. Future theoretical and experimental studies may find useful hints from the analysis herein.

\begin{acknowledgments}
The author acknowledges Rahul Nandkishore and J\"orn Venderbos for fruitful discussions and feedback on the manuscript, and especially thanks Rahul Nandkishore for encouragement on this work. This research was sponsored by the Army Research Office and was accomplished under Grant No. W911NF-17-1-0482. The views and conclusions contained in this document are those of the authors and should not be interpreted as representing the official policies, either expressed or implied, of the Army Research Office or the U.S. Government. The U.S. Government is authorized to reproduce and distribute reprints for Government purposes notwithstanding any copyright notation herein.
\end{acknowledgments}




\bibliography{Reference}

\end{document}